\documentclass[useAMS,usenatbib]{mn2e}
\usepackage{graphicx}
\usepackage{hyperref}
\usepackage[all]{hypcap}

\include{journal_commands}

\title[CMB Power Spectrum Likelihood with ILC]{CMB Power Spectrum Likelihood with ILC}
\author[Jason Dick, Guillaume Castex, Jacques Delabrouille]{Jason
  Dick$^1$\thanks{E-mail: dick@sissa.it}, Guillaume Castex$^{2}$\thanks{E-mail: castex@apc.univ-paris7.fr}, Jacques Delabrouille$^2$\thanks{E-mail: delabrouille@apc.univ-paris7.fr}\\
$^{1}$ SISSA,
via Bonomea 265,
34136 Trieste,
Italy\\
$^2$ APC,
10, rue Alice Domon et L\'eonie Duquet,
75205 Paris Cedex 13,
France}
\begin{document}

\date{Submitted to MNRAS 21/3/2012}

\pagerange{\pageref{firstpage}--\pageref{lastpage}} \pubyear{2012}

\maketitle

\label{firstpage}

\begin{abstract}
We extend the ILC method in harmonic space to include the error in its CMB
estimate.  This allows parameter estimation routines to take into account the
effect of the foregrounds as well as the errors in their subtraction in
conjunction with the ILC method.  Our method requires the use of a model of
the foregrounds which we do not develop here.  The reduction of the foreground
level makes this method less sensitive to unaccounted for errors in the
foreground model.  Simulations are used to validate the calculations and
approximations used in generating this likelihood function.
\end{abstract}


\section{Introduction}

The Cosmic Microwave Background (CMB) was emitted when our universe was
approximately 380,000 years old, and provides a window into the physics of our
early universe.  In particular, most of the information about early physics is
encoded in the angular power spectrum of the temperature anisotropies
\citep[e.g.][]{hu02e}, for the reason that all of the information of a
statistically isotropic Gaussian random field is encoded in its angular power
spectrum, and the deviations of the CMB anisotropies from this are at most
small, as estimates of the primordial non-Gaussianity of the CMB are
consistent with zero as measured by WMAP \citep{komatsu09, komatsu11}.  While
the CMB we observe is known to not be exactly Gaussian due to the
gravitational lensing of the intervening matter \citep{zaldarriaga99,
  okamoto03, lewis06}, the power spectrum remains a powerful tool for
extracting information about early physics \citep{lewis05}.  One difficulty is
that there are a number of sources between us and the CMB that emit radiation
in the same frequency range.  A commonly-used method for subtracting these
foregrounds from the CMB is known as Internal Linear Combination (ILC).  This
method was first described in \citet{tegmark98}.  It is internal in the sense
that it only depends upon data related to the experiment at hand.  And it is
linear in that the estimate of the CMB is computed as a linear combination of
the observed temperature maps, subject to the constraint that the sum of the
linear weights is equal to one.  This constraint ensures that with the maps
calibrated in thermodynamic units with the CMB emission being the same in all
frequencies, the resultant estimate of the CMB has unit response to the CMB.

In \citet{bennet03a}, a nonlinear fitting routine was used to compute the set
of linear weights which minimized the variance of the output map.  Later,
\citet{eriksen04a} showed that the ILC result could be computed analytically
by the use of a Lagrange multiplier, as shown previously in \citet{tegmark98}.
The WMAP team later extended their ILC method in \citet{hinshaw07} to include
an attempt to remove the bias from the ILC method through a series of Monte
Carlo simulations.  They detected only a small bias in these simulations.
However, we argue in Appendix \ref{app:bootstrap} that this method may
severely underestimate the ILC bias.  They make no mention of incorporating
any error in their ILC estimate, and as a result avoid using the ILC map for
most of their cosmological studies, instead opting for a template fitting
method whose noise properties are more easily-understood.  They do make use of
their ILC map for their low-$\ell$ likelihood \citep{hinshaw09, larson11}, but
because they perform the ILC in pixel space, and because they do not compute
the ILC weights in the same region of the sky where they apply them, they are
unlikely to suffer problems related to the ILC bias.  Furthermore, as they are
strongly dominated by cosmic variance errors for their low-$\ell$ analysis,
their lack of detailed consideration of the noise properties of their ILC map
is unlikely to undermine their analysis.

Extending the pixel-based analysis performed in \citet{bennet03a},
\citet{tegmark03} estimated the weights independently on subdivisions in both
pixel space and in harmonic space to produce a high-resolution CMB map.  For
validation, they demonstrate that their results are similar to the WMAP team's
in \mbox{\citet{bennet03a}}.  A similar analysis was performed in
\citet{delabrouille09}; \citet{basak12} where the subdivision of the sky was
performed on needlets, which are localized in both pixel space and harmonic
space.  They included a calculation of the ILC bias that results from the fact
that in minimizing the total variance of the output CMB estimate, the ILC
method can exploit chance correlations between the CMB and the foregrounds to
partially cancel the CMB.  However, this calculation assumes that the needlet
coefficients are independent, which is not exactly the case.  Validation of
their results were performed using Planck Sky Model simulations.

If we are working with the full-sky, however, the harmonic coefficients are
independent.  Taking advantage of this fact, a harmonic-space analysis was
examined in \citet{saha08} who compute a similar bias, while using simulations
to estimate the bias and the error in the bias.  Meanwhile, \citet{kim08} also
investigated ILC in harmonic space, and made use of an iterative method in an
attempt to reduce the foreground bias.  However, this iterative method is
likely to be susceptible to the same pitfall that we describe in Appendix
\ref{app:bootstrap}, because their estimate of the foregrounds is simply the
data minus their ILC estimate of the CMB.

Moving beyond temperature analysis, \citet{amblard07} made use of the ILC
technique to subtract the foregrounds in harmonic space for use in measuring
the $E$ and $B$-mode polarization of the CMB, while \citet{efstathiou09}
argued that the bias induced by the ILC method was too large to allow for
estimation of the CMB $B$-mode signal, instead opting for a template-based
parametric foreground removal method.  We do not consider polarization here,
as though all of our calculations would remain unchanged for the $E$ and
$B$-mode power spectrum estimates, the effect of ILC on the $TE$ cross
spectrum is far more difficult, and polarization has a much larger foreground
signal compared to the CMB than is the case with temperature data, suggesting a
dedicated analysis would be preferred.

The previous methods have all made use of simulations for validation, with the
furthest any has gone to compute the errors being to estimate the variance of
a series of simulation results.  Ideally, we would like to have a way to
compute the errors in the ILC subtraction method, or at least a method for
propagating the errors in a foreground model.  To this end, we describe the
derivation of our approximation to the true likelihood that a given
theoretical model matches the harmonic-space ILC output in Section
\ref{sec:likelihood}.  This is followed by our validation procedures in
Section \ref{sec:validation} and a discussion of what is required to make use
of this likelihood function in a realistic experiment in Section
\ref{sec:discussion}.

\section[Likelihood Derivation]{Derivation of the ILC power spectrum likelihood}
\label{sec:likelihood}

In this paper, we follow the same harmonic-space Internal Linear Combination
(ILC) analysis used in \citet{saha08}.  Performing the analysis in harmonic
space allows us to take advantage of the independence of the harmonic
coefficients of the CMB to simplify the calculation.  Their primary result was
an estimate of the bias in the power spectrum of the estimated CMB map induced
by the ILC method, which we reproduce here:
\begin{equation}
\langle \tilde{C}_\ell \rangle = \langle C_\ell \rangle + {1 - n_c \over
  2\ell+1}\langle C_\ell \rangle + {1 \over \bld{e}^\dagger
  \hat{\mathbf{N}}_\ell^{-1} \bld{e}}.
\label{eqn:bias}
\end{equation}
Here $\langle \tilde{C}_\ell \rangle$ is the expectation value of the power
spectrum of the ILC-estimated CMB map, $C_\ell$ is the true
power spectrum of the CMB, $n_c$ is the number of channels, $\bld{e}$ is a
vector of all ones, $\hat{\mathbf{N}}_\ell$ is the channel-channel covariance
matrix of the foregrounds plus noise at each $\ell$, and $^\dagger$ is the
conjugate transpose.  Following \citet{saha08}, $\hat{\mathbf{N}}_\ell$ is
assumed to be fixed and known, with the expectation value only over CMB
realizations.  We discuss how to relax this assumption later in Section
\ref{sec:discussion}.

We can understand the three main terms of this equations as follows.  The
first term is simply the power spectrum of the CMB, which comes out with a
factor of unity because of the assumption of the frequency scaling of the CMB
signal combined with the assumption of perfect calibration.  As shown in
\citet{dick10}, calibration error that is significant compared to the ratio of
noise to signal can result in pathological behavior of the ILC method.  We
assume in this paper that the data is of sufficient quality that such
pathological behavior is avoided, though it may be worth investigating later
whether or not this is a valid assumption, and whether our likelihood analysis
can be modified to include this possibility.

The second term in the equation is the bias due to chance correlations between
the CMB and the foregrounds.  It is simply proportional to the power spectrum
of the CMB because the Gaussianity of the CMB allows us to ignore many of
the properties of the foregrounds entirely, such as their spatial
correlations.  The only assumption here is a full-rank foreground
covariance (this assumption breaks at very low $\ell$, if $2\ell+1 < n_c$,
but can be recovered by binning the power spectrum).

The final term is a contamination term: it is the power spectrum of the ILC
method performed on the foreground plus noise signal alone.  The magnitude of
this term depends upon the frequency coverage of the instrument: higher
frequency coverage allows for better removal of the foregrounds, reducing the
magnitude of this contamination.

In the following subsection, we extend this calculation to an estimate of the
variance on the power spectrum, which allows us to develop an approximation to
the full probability distribution.

\subsection{Variance of the estimated power spectrum}
The variance of the ILC estimate of the CMB can be computed using the same
general method as used in computing the bias.  However, we differ
substantially in the specific calculation from that described in
\citet{saha08}.  In particular, they make the assumption that the foreground
covariance is not full rank.  Because of the instrument noise ensuring that
this covariance is indeed full rank, we do not make this assumption.  So first
we describe the much shorter calculation of the ILC bias, then present an
outline of the calculation for the variance of the ILC estimate.  A fuller
description of the calculation can be found in Appendix \ref{app:variance}.

First, we start with a few definitions.  As the calculations are performed in
harmonic space, $a_{\ell m}$ is the spherical harmonic transform of the true
CMB, and we define $\bld{f}_{\ell m}$ as the spherical harmonic transform of
everything except the CMB, which includes foregrounds and instrument noise.
As the foreground plus noise signal in the sky varies from frequency to
frequency, this latter term is a vector for each $\ell$, $m$ pair.  The
spherical harmonic transform of the observed sky, then, can be written as:
\begin{equation}
\bld{d}_{\ell m} = \bld{e} a_{\ell m} + \bld{f}_{\ell m}.
\label{eqn:mapdef}
\end{equation}
Here $\bld{d}_{\ell m}$ is a vector representing the spherical harmonic
transforms of the observations at each frequency.  The ILC method selects a
series of linear weights $\bld{w}$ for each channel which minimizes the
variance of the output subject to the constraint that $\bld{w}^\dagger\bld{e}
= 1$.  This constraint forces the contribution of the CMB in the final result
to be unity under the assumption that the maps are accurately calibrated in
thermodynamic units (where the CMB anisotropies are identical in each channel).
Minimizing the variance of the estimated CMB $\tilde{a}_{\ell m}$ results
in:
\begin{equation}
\tilde{a}_{\ell m} = \bld{w}^\dagger \bld{d}_{\ell m},
\end{equation}
\begin{equation}
\bld{w}^\dagger = {\bld{e}^\dagger \hat{\mathbf{C}}_\ell^{-1} \over \bld{e}^\dagger
  \hat{\mathbf{C}}_\ell^{-1} \bld{e}}.
\end{equation}
Here $\hat{\mathbf{C}}_\ell$ is the empirical channel-channel covariance of the
observations at each $\ell$, defined as:
\begin{equation}
\hat{\mathbf{C}}_\ell = {1 \over 2\ell + 1} \sum_{m=-\ell}^{\ell} \bld{d}_{\ell m}
\bld{d}^\dagger_{\ell m}.
\end{equation}
The fact that $\bld{d}_{\ell m}$ is the spherical harmonic transform of a
set of real-valued maps on the sphere ensures the matrix
$\hat{\mathbf{C}}_\ell$ only has real-valued elements.  This becomes in the
computation of the variance.

With these definitions, it is easy to show that the power spectrum of the
recovered map takes on the simple form:
\begin{equation}
\tilde{C}_\ell = {1 \over \bld{e}^\dagger \hat{\mathbf{C}}_\ell^{-1} \bld{e}}.
\label{eqn:powspec_est}
\end{equation}

From here we can compute how the resultant power spectrum depends upon the
true CMB and the foregrounds plus noise.  This is done by substituting Equation
(\ref{eqn:mapdef}) into the expression for the covariance, giving:
\begin{equation}
\hat{\mathbf{C}}_\ell = \sum_{m=-\ell}^{\ell} \left(a_{\ell m}a^*_{\ell m}
\bld{e e}^\dagger + a_{\ell m} \bld{e f}^\dagger_{\ell m} + \bld{f}_{\ell m}
\bld{e}^\dagger a^*_{\ell m} + \bld{f}_{\ell m} \bld{f}_{\ell
  m}^\dagger\right).
\label{eqn:covariance_long}
\end{equation}

The first term is simply the power spectrum of the true CMB signal.  The last
term is the covariance of the foregrounds plus noise.  The two terms in the
middle are due to the chance correlation between the CMB and foregrounds plus
noise.  We can simplify this equation using notation similar to that in
\citet{saha08} by defining:
\begin{equation}
\bld{f}_\ell \equiv \sum_{m=-\ell}^{\ell} \bld{f}_{\ell m} a^*_{\ell m}.
\label{eqn:fl_def}
\end{equation}

As before, the fact that both $\bld{f}$ and $a$ are spherical harmonic
transforms of real-valued maps ensures that the chance correlation between the
CMB and foregrounds plus noise $\bld{f}_\ell$ is also real-valued.
Equation (\ref{eqn:covariance_long}) can then be written as:
\begin{equation}
\hat{\mathbf{C}}_\ell = C_\ell \bld{e e}^\dagger + \bld{e f}^\dagger_\ell +
\bld{f}_\ell\bld{e}^\dagger + \hat{\mathbf{N}}_\ell.
\label{eqn:covariance_short}
\end{equation}
Here we have introduced $\hat{\mathbf{N}}_\ell$ to represent the covariance
of the foregrounds plus noise.  If we combine this equation with the identity
$(\mathbf{A} + \bld{b c}^\dagger)^{-1} = \mathbf{A}^{-1} - \lambda^{-1}
\mathbf{A}^{-1}\bld{b c}^\dagger\mathbf{A}^{-1}$, where $\lambda = 1 +
\bld{c}^\dagger\mathbf{A}^{-1}\bld{b}$, we can obtain an analytical form for
the power spectrum of the ILC estimate of the CMB.

The use of this identity requires that both $\hat{\mathbf{C}}_\ell$ and
$\hat{\mathbf{N}}_\ell$ be invertible, which also implies that $\lambda \neq
0$.  This is the case for instrument noise alone, and thus is the case for
foregrounds plus instrument noise.  After applying this identity to the inverse
of the covariance of the observations three times and substituting the result into
Equation (\ref{eqn:powspec_est}), we obtain:
\begin{eqnarray}
\tilde{C}_\ell &=& C_\ell \nonumber\\
&&+ {1 + \bld{e}^\dagger \hat{\mathbf{N}}_\ell^{-1}
  \bld{f}_\ell + \bld{f}^\dagger_\ell \hat{\mathbf{N}}_\ell^{-1} \bld{e}
  + \bld{e}^\dagger \hat{\mathbf{N}}_\ell^{-1} \bld{f}_\ell\bld{f}^\dagger_\ell
  \hat{\mathbf{N}}_\ell^{-1} \bld{e} \over \bld{e}^\dagger
  \hat{\mathbf{N}}_\ell^{-1} \bld{e}} \nonumber\\
 &&- \bld{f}^\dagger_\ell \hat{\mathbf{N}}_\ell^{-1} \bld{f}_\ell.
\label{eqn:est_cl}
\end{eqnarray}
Making use of the independence of the $a_{\ell m}$ values, with $\langle a_{\ell
  m_1} a^*_{\ell m_2}\rangle = \delta_{m_1 m_2} C_\ell$, it is easy to show that
the expectation value of the above equation is Equation (\ref{eqn:bias}).

\subsection{Variance of the bias}

While we leave a fuller explanation of the computation of the variance in the
ILC bias to Appendix \ref{app:variance}, here we sketch an outline of the
general process.  As a first step in the calculation, we subtract the
expectation value $\langle \tilde{C}_\ell \rangle$ from the estimate
$\tilde{C}_\ell$:
\begin{eqnarray}
\lefteqn{\tilde{C}_\ell - \langle \tilde{C}_\ell \rangle =} \nonumber\\
 && C_\ell - \bld{f}^\dagger_\ell \hat{\mathbf{N}}_\ell^{-1}
  \bld{f}_\ell\nonumber\\
 && + {1 + \bld{e}^\dagger \hat{\mathbf{N}}_\ell^{-1}
  \bld{f}_\ell + \bld{f}^\dagger_\ell \hat{\mathbf{N}}_\ell^{-1} \bld{e}
  + \bld{e}^\dagger \hat{\mathbf{N}}_\ell^{-1} \bld{f}_\ell\bld{f}^\dagger_\ell
  \hat{\mathbf{N}}_\ell^{-1} \bld{e} \over \bld{e}^\dagger
  \hat{\mathbf{N}}_\ell^{-1} \bld{e}}\nonumber\\
 && - \langle C_\ell \rangle - {1 - n_c \over
  2\ell+1}\langle C_\ell \rangle - {1 \over \bld{e}^\dagger
  \hat{\mathbf{N}}_\ell^{-1} \bld{e}}.
\end{eqnarray}
This allows us to group $C_\ell$ with its expectation value, and cancel
$1/\bld{e}^\dagger \hat{\mathbf{N}}_\ell^{-1} \bld{e}$, leaving us with an
equation with no more terms than in Equation (\ref{eqn:est_cl}).  This gives us:
\begin{eqnarray}
\lefteqn{\delta\tilde{C}_\ell =} \nonumber\\
 && \delta C_\ell - \bld{f}^\dagger_\ell \hat{\mathbf{N}}_\ell^{-1}
  \bld{f}_\ell - {1 - n_c \over
  2\ell+1}\langle C_\ell \rangle\nonumber\\
 && + {\bld{e}^\dagger \hat{\mathbf{N}}_\ell^{-1}
  \bld{f}_\ell + \bld{f}^\dagger_\ell \hat{\mathbf{N}}_\ell^{-1} \bld{e}
  + \bld{e}^\dagger \hat{\mathbf{N}}_\ell^{-1} \bld{f}_\ell\bld{f}^\dagger_\ell
  \hat{\mathbf{N}}_\ell^{-1} \bld{e} \over \bld{e}^\dagger
  \hat{\mathbf{N}}_\ell^{-1} \bld{e}}.
  \label{eqn:delta_cl}
\end{eqnarray}

The next step involves squaring Equation (\ref{eqn:delta_cl}) and taking the
expectation value, the full calculation of which is relegated to Appendix
\ref{app:variance}.  Here we simply note that the square of Equation
(\ref{eqn:delta_cl}) can be considered in two groups: parts that have even
powers of the $a_{\ell m}$ values, and parts that have odd powers of the
$a_{\ell m}$ values.  When we take the expectation value, only those
components that have even powers in the $a_{\ell m}$ values survive, so we can
immediately throw out a large number of the cross-terms in the square, making
this rather cumbersome equation more manageable. The end result is the
following:
\begin{equation}
\langle \delta\tilde{C}_\ell^2 \rangle = {2 \langle C_\ell \rangle \over
  2\ell+1}\left(\langle C_\ell \rangle + {1 - n_c \over 2\ell+1}\langle C_\ell
\rangle + {2 \over \bld{e}^\dagger \hat{\mathbf{N}}_\ell^{-1} \bld{e}}\right).
\label{eqn:variance}
\end{equation}

Equation (\ref{eqn:variance}) is the main result of our paper and can be
broken down as follows.  The first term is simply the standard cosmic variance
term.  The second term is the modification of the cosmic variance term that
stems from the existence of the ILC bias.  There is no term related to the
contamination of the ILC result by the foregrounds plus noise, because that
term is assumed to be fixed.  The final term proportional to
$2/\bld{e}^\dagger \hat{\mathbf{N}}_\ell^{-1} \bld{e}$ instead comes
from the square of the correlation term $(\bld{e}^\dagger
\hat{\mathbf{N}}_\ell^{-1} \bld{f}_\ell + \bld{f}^\dagger_\ell
\hat{\mathbf{N}}_\ell^{-1} \bld{e}) /\bld{e}^\dagger
\hat{\mathbf{N}}_\ell^{-1} \bld{e}$, indicating that this represents the
variance of the zero-mean component of the chance correlations between the CMB
and foregrounds.

It should be noted at this point that the variation in the noise, as well as
our uncertainties in the foreground model itself, have not yet been taken into
account.  Instead, we are providing a likelihood function which is a function
of the theoretical power spectrum and a specific realization of the
foregrounds and noise.  The uncertainty in the foreground plus noise signal
must be taken into account separately.  Minimal modifications of existing
parameter estimation pipelines may be ideal for taking into account the fact
that the foregrounds plus noise are not perfectly-known.  We discuss this in
more detail in Section \ref{sec:discussion}.

For now, however, we produce a modification of Equation (\ref{eqn:variance})
making the assumption that the foregrounds plus noise are
Gaussian-distributed.  This results in the addition of a cosmic variance-like
term, but making use of the foreground power spectrum instead:
\begin{eqnarray}
\langle \delta\tilde{C}_\ell^2 \rangle &=& {2 \langle C_\ell \rangle \over
  2\ell+1}\left(\langle C_\ell \rangle + {1 - n_c \over 2\ell+1}\langle C_\ell
  \rangle + {2 \over \bld{e}^\dagger \hat{\mathbf{N}}_\ell^{-1}
  \bld{e}}\right)\nonumber\\
  &&+ {2 \over 2\ell+1}\left({1 \over \bld{e}^\dagger \hat{\mathbf{N}}_\ell^{-1}
  \bld{e}}\right)^2.
\label{eqn:mod_var}
\end{eqnarray}
While this approximation is useful for obtaining a rough estimate of the total
error in the extracted CMB, we would not recommend using it for a serious
analysis without detailed verification due to its ad-hoc nature.

\subsection{Approximating the full probability distribution}
The previous analysis indicates that we can estimate the full probability
distribution of the power spectrum that results from the ILC method given only
a model of the true power spectrum and a model of the covariance of the
foregrounds, which would allow the use of these calculations in CMB parameter
estimation.  To do this, we go back and examine each term in the mean and
variance of the ILC-estimated CMB sky.

In particular, when we square $\delta \tilde{C}_\ell$ and take the expectation
value, we are left with terms which stem from the fourth power of the
spherical harmonic transform coefficients ($a_{\ell m}$) and terms which stem
from the second power of the spherical harmonic transform coefficients.  The
fourth power terms become the first two terms in Equation
(\ref{eqn:variance}), and the second power terms become the third.

We can surmise that the fourth power terms collectively act as a modified
$\chi^2$ distribution in $\tilde{C}_\ell$, while the second power terms act as
a Gaussian, and we can estimate the parameters of these respective
distributions by comparing the mean and variance of these two terms.

For a $\chi^2$ distribution with $k$ degrees of freedom, the mean and variance
are:
\begin{eqnarray}
\langle \chi_k^2 \rangle &=& k, \\
\langle (\chi_k^2 - \langle \chi_k^2 \rangle)^2\rangle &=& 2k.
\end{eqnarray}
The distribution of $C_\ell$ differs from this distribution by a factor of
$\langle C_\ell \rangle/k$, with $k = 2\ell+1$.  We can thus parameterize this
distribution with two parameters, $\sigma^2 = \langle C_\ell \rangle$ and $k
= 2\ell+1$.  The mean and variance of $C_\ell$ then becomes:
\begin{eqnarray}
\langle C_\ell \rangle &=& \sigma^2, \\
\langle (C_\ell - \langle C_\ell \rangle)^2 \rangle &=& {2 \sigma^2 \over k}.
\end{eqnarray}
Similarly, we can define new parameters appropriate for the first two terms in
the mean and variance of $\tilde{C}_\ell$:
\begin{eqnarray}
\tilde{\sigma}^2 &=& \left(1 + {1-n_c \over 2\ell+1}\right)\langle C_\ell \rangle, \\
\tilde{k} &=& \left(1 + {1-n_c \over 2\ell+1}\right) (2\ell+1).
\end{eqnarray}

The remaining terms in the mean and sigma can be modeled as a constant offset
($1/\bld{e}^\dagger \hat{\mathbf{N}}_\ell^{-1} \bld{e}$) and a zero-mean
Gaussian with variance equal to $4\langle C_\ell \rangle
/((2\ell+1)\bld{e}^\dagger \hat{\mathbf{N}}_\ell^{-1} \bld{e})$.  Therefore
our approximation to the full probability distribution of $\tilde{C}_\ell$ is
modeled as the sum of two random variables, one following a Gaussian
distribution and the other following a modified $\chi^2$ distribution.

\subsection{Likelihood evaluation}
In order to estimate the likelihood of a cosmological model given the data, we
need to write down the full probability distribution of the sum of a modified
$\chi^2$ distribution and a Gaussian.  To examine this, we consider the
random variable $x$ which has a probability distribution $P_x(x)$, the
random variable $y$ which has a probability distribution $P_y(y)$, and their
sum $s$ which has a probability distribution $P_s(s)$.  With these
definitions, it is easy to see that the probability of the sum $s$ is:
\begin{equation}
P_s(s) = \int_{-\infty}^{\infty}\int_{-\infty}^{\infty} P_x(x)P_y(y)\delta(s - x - y) dxdy.
\end{equation}
From this, we can write the unnormalized probability distribution of the
estimated $\tilde{C}_\ell$ as:
\begin{equation}
P(\tilde{C}_\ell|C_\ell, \hat{\mathbf{N}}_\ell) \propto \int_0^{\infty}
x^{{\tilde{k} \over 2} - 1} e^{-{x \over 2\tilde{\sigma}_k^2} -
  {(\tilde{C}_\ell - x)^2 \over 2 \sigma_G^2}} dx,
\label{eqn:likelihood}
\end{equation}
where $\tilde{\sigma}_k^2$ is the variance of the modified $\chi^2$
distribution, and $\sigma_G^2$ is the variance of the Gaussian distribution.
This integral is easily computed numerically.

\section{Validation}
\label{sec:validation}

In order to validate this likelihood approximation, we perform a series of
simulations.  The first set of simulations is performed merely to verify that
our calculations were accurate, using $10^5$ simulations at a single multipole
moment with a toy model of the foregrounds.  The second simulation is a full
Planck Sky Model simulation where we could verify that this likelihood
estimate holds up under a more realistic scenario, and further provides some
insight into the likely relative magnitudes of the various terms in the
likelihood for real data.

\subsection{Toy simulations}

For toy simulations, we use a minimalistic mathematical model purely in order
to verify the calculations in this paper were performed accurately.  To
this end, we take the toy CMB to be a set of unit-variance independent complex
Gaussian random variables (the $a_{\ell m}$ values), constrained so that $a_{\ell
  m} = a^*_{\ell -m}$.  To ensure numerical accuracy, we ran the simulation
over one million realizations of these Gaussian random variables for a single
$\ell$.

\begin{figure}
\begin{center}
\includegraphics[width=0.5\textwidth]{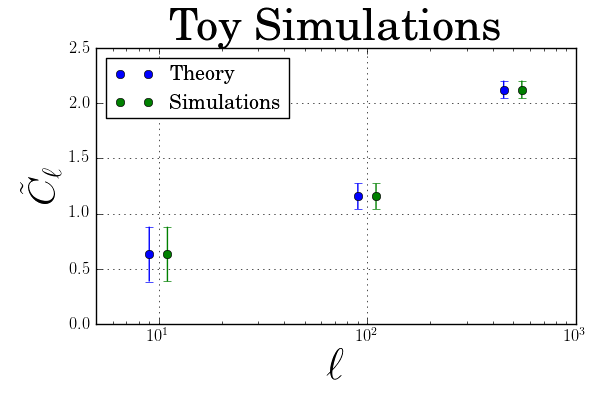}
\end{center}
\caption{The mean and error bars given by the square root of the variance of a
  set of $10^5$ simulations are compared against the expected results in
  Equations (\ref{eqn:bias}) and (\ref{eqn:variance}).  The differences are
  indistinguishable in the plot, and numerically differ by less than 0.3\%,
  the expected deviation from $10^5$ simulations.  In using test data that
  precisely match the assumptions used in deriving the mean and variance,
  these toy simulations verify that Equations (\ref{eqn:bias}) and
  (\ref{eqn:variance}) were derived correctly.}
\label{fig:toysims}
\end{figure}

For the foregrounds plus noise, since the calculations assumed a fixed
foreground plus noise signal, we generate one set of correlated complex
Gaussian random variables for each of nine channels.  The correlations are
produced by making use of a randomly-generated covariance matrix for each of
the nine channels (the square of a random symmetric matrix with unit variance
elements).  This covariance matrix is then divided by a factor of 100 in order
to make it relatively small compared to the variance of the toy CMB.  The mean
and variance of these results for three different choices of $\ell$ are shown
in Figure \ref{fig:toysims}.

\begin{figure*}
\begin{center}
\includegraphics[width=0.33\textwidth]{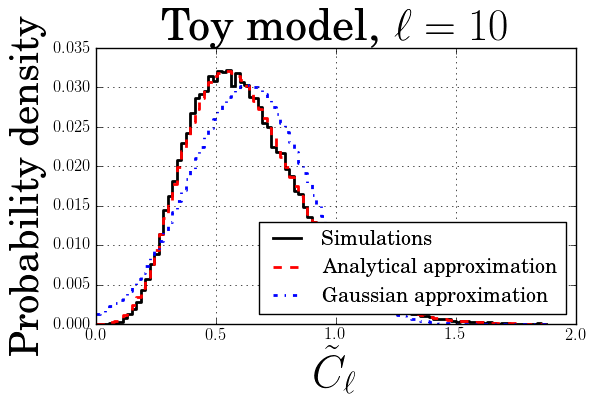}
\includegraphics[width=0.33\textwidth]{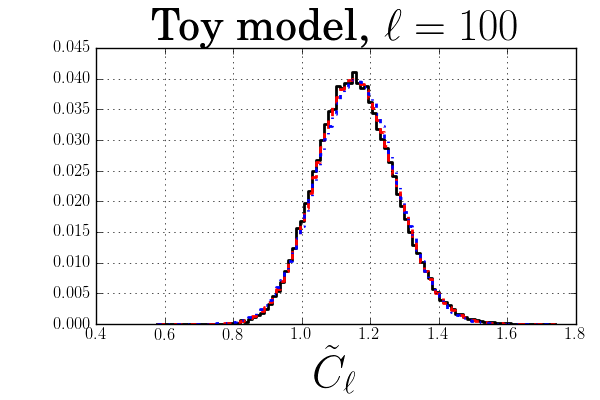}
\includegraphics[width=0.33\textwidth]{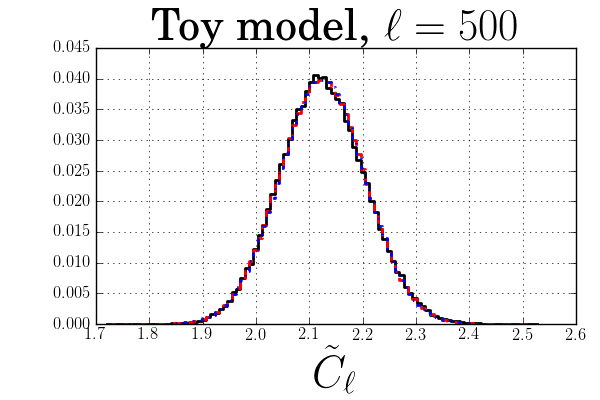}
\end{center}
\caption{This test of the approximations to the probability distribution of
  $\tilde{C}_\ell$ shows that the analytical approximation, which models
  the probability as a sum of a chi-square distributed random variable and a
  Gaussian-distributed random variable, as well as a Gaussian approximation
  for comparison.  While not perfect, the analytical approximation is
  better than the Gaussian at replicating the simulations.  At high $\ell$,
  the Gaussian approximation and the analytical approximation become
  nearly indistinguishable in these plots.}
\label{fig:toy}
\end{figure*}

Secondly, in order to validate the analytical approximation to the full
probability distribution, we bin the results of the above simulations and
compare them against the integral of the probability distribution across the
bin (Figure \ref{fig:toy}).  In each case, the analytical approximation to the
full probability distribution matches the simulations very well.  The
correspondence is not expected to be perfect.  However, the analytical
approximation is clearly superior to the Gaussian approximation at low to
medium $\ell$, with the two becoming nearly indistinguishable at $\ell$ values
of around a few hundred.

\subsection{Analysis of realistic simulations}

In order to verify our likelihood function in a more realistic situation, we
use a single realization of the foregrounds at all nine Planck frequency
channels using the Planck Sky Model (described in Section
\ref{sec:psm}).  To the simulated foregrounds, we add a single realization of
the noise.  This realization is computed as uncorrelated, non-uniform noise,
utilizing the same hit maps and per-sample variances as in \citet{leach08}.
To the simulated sky and noise we add a simulated lensed CMB using the LensPix
software package \citep{lewis05}.  A lensed CMB realization is used for the
reason that the lensing breaks the assumption of complete independence of the
$a_{\ell m}$ values, and is known to be a real feature in the observed CMB.
Taking this non-ideality into account is thus important to help ensure that
our method will be applicable to real data.

\subsubsection{Planck Sky Model}
\label{sec:psm}
Sky maps used in the present work are based on simulations obtained with the
pre-launch Planck Sky Model
software\footnote{\url{http://www.apc.univ-paris7.fr/~delabrou/PSM/psm.html}},
a code designed to generate realistic full-sky maps of sky emission at
millimetre wavelengths.  A complete description of the PSM code is given in
\citet{delabrouille12}.  We summarize here the main aspects of the particular
simulation we use.  To make a realization of the foregrounds, the PSM
generates maps of different diffuse and compact source components, which are
then coadded. For most of them, realizations are based on observed data sets,
complemented by theoretical modelling where and when observations are
missing. The map for each component is then computed at a set of observation
frequencies on the basis of a modelled emission law for the component.  In
addition to CMB anisotropies, the components of sky emission implemented in
the PSM include point sources (galactic and extragalactic), Sunyaev-Zel'dovich
effects (thermal and kinetic), diffuse galactic emission, and fluctuations of
the far infrared background emission from high redshift galaxies.

In the present simulations, diffuse emission from the galactic interstellar
medium (ISM) is modelled as the sum of four main components: thermal emission
of small dust particles heated by radiation from nearby stars; synchrotron
emission, due to relativistic electrons spiralling in the galactic magnetic
field; Free-free emission from warm regions of ionised interstellar gas; and
the emission of small rotating dust grains.

The thermal dust emission is modelled on the basis of model 7 of
\citet{finkbeiner99}. The total emission is represented as the sum of two
modified blackbodies of the form:
\begin{equation}
I_{\nu} = \sum_{i=1}^{2} N_{i}\epsilon_{i}\nu^{\beta_{i}}B_{\nu}(T_{i})
\end{equation}
where $B_{\nu}(T_i)$ is the Planck function at temperature $T_i$, $N_i$ is the
column density of species $i$, and $\epsilon_{i}\nu^{\beta_{i}}$ accounts for
the normalization and frequency dependence of the emissivity.

Low frequency diffuse foregrounds (synchrotron, free-free ad spinning dust)
are based on the analysis of WMAP observations performed by
\citet{miville-deschenes08}.  The synchrotron emission is obtained by
scaling with a pixel-dependent emission law the template emission map observed
at 408 MHz by \citet{haslam82}. In the radio-mm frequency range, for an
electron density following a power law of index $p$, ($ne(E) \propto E−p$),
the synchrotron frequency dependence is also modelled as a power law:
\begin{equation}
I_{sync}(\nu) \propto \nu^{\beta_s+2}
\end{equation}
where the spectral index is equal to $\beta_{s} = -(p + 3)/2$.  The
synchrotron spectral index depends on cosmic ray properties. It varies with
the direction on the sky, and possibly, with the frequency of observation
(see, e.g., \citet{strong07}, for a review of propagation
and interaction processes of cosmic rays in the Galaxy).  The template
synchrotron map obtained from Haslam et al., corrected for an offset monopole
of $8.33 K$ (Rayleigh-Jeans), is extrapolated in frequency on the basis of the
spectral index map obtained from WMAP data.

Free-free is modelled on the basis of a template free-free map at 23 GHz, and
a single universal emission law as given by \citet{dickinson03}.

In the PSM, the spinning dust emission is represented on the basis of a single
template, together with a unique emission law which can be parametrised
following the model of \citet{draine98}. The spinnning dust template
is obtained by removing every other components contributing to the WMAP 23GHz
data (free-free, synchrotron and thermal dust).

As the resolution of our simulation ($5$ arcmin) is better than that of the
templates used to construct the model, small-scale fluctuations are added to
Galactic emission simulations. The method used is described in
\citet{miville-deschenes07}.  A Gaussian random field $G_{ss}$ having
a power spectrum defined as:
\begin{equation}
C_{l} = l^{\gamma}\left[ e^{-l^{2}\sigma^2_\mathrm{sim}} -e^{l^2\sigma^2_\mathrm{tem}}\right]
\end{equation}
is generated. Here $\sigma_\mathrm{sim}$ and $\sigma_\mathrm{tem}$ are the resolutions (in
radian) of the simulation ($5'$ here) and of the template to which small-scale
fluctuations are added. The zero mean Gaussian field $G_{ss}$ is then
normalized and multiplied by the template map exponentiated to a power $\beta$
in order to generate the proper amount of small-scale fluctuations as a
function of local intensity.  The resulting template map with small scale
fluctuation added is then:
\begin{equation}
I^\prime_\mathrm{tem} = I_\mathrm{tem} + \alpha G_{ss} I^{\beta}_\mathrm{tem}
\end{equation}
where $\alpha$ and $\beta$ are estimated for each template in order to make
sure that the power spectrum of the small-scale structure added is in the
continuity of the large scale part of the template.

The Sunyaev-Zel'dovich effect is modelled as the sum of two effects: thermal
SZ effect which corresponds to the interaction of the CMB with a hot,
thermalised electron gas, and the kinetic SZ effect which corresponds to CMB
interaction with electrons having a net ensemble peculiar velocity along the
line of sight.  For thermal SZ, in the approximation of non-relativistic
electrons, the change in sky brightness is:
\begin{equation}
\delta I_\nu = yf(\nu)B_\nu(T_\mathrm{CMB})
\end{equation}
where $B_ν(T_\mathrm{CMB})$ is the CMB blackbody spectrum, $f ( \nu )$ is a universal
function of frequency that does not depend on the physical parameters of the
electron population and $y$, the Compton parameter, is proportional to the
integral along the line of sight of the electronic density $ne$ multiplied by
the temperature of the electron gas:
\begin{equation}
\delta y = \int \frac{kT_e}{m_ec^2}n_e\sigma_Tdl
\end{equation}
where $T_e$ is the temperature of the electron gas, characterized by a Fermi
distribution.  The kinetic effect, due to the interaction of CMB photons with
moving electrons, results in a shift of the photon distribution seen by an
observer on Earth:
\begin{equation}
\delta I_\nu = -\beta_r \tau \left[ \frac{\partial B_\nu (T)}{\partial T}\right]_{T=T_\mathrm{CMB}}
\end{equation}
where $\beta_r$ is the dimensionless cluster velocity along the line of sight.
The SZ simulation used in the present work is based on two different parts:
hydro+N-body simulations of the distribution of baryons for redshifts $z <
0.025$ obtained from \citet{dolag05}, and pure N-body simulations of dark
matter structures in a Hubble volume, using template from a smaller simulation
by \citet{schafer06}.

Other compact sources are implemented as the sum of three populations: radio
sources, infrared sources, and galactic ultra-compact HII regions.

Radio sources are simulated on the basis of both real radiosources observed
around 1 and/or 5 GHz and on simulated sources generated at random to
homogeneize the coverage where the sky surveys are shallow or incomplete.
Sources fluxes are extrapolated to all frequencies required by the simulation
using power law approximations of the spectra of the sources, of the form
$S_\nu \propto \nu^{-\alpha}$ with different spectral indices in different
frequency domains.  Radio sources are divided in two populations, steep and
flat, with typical spectral indexes distributed according to a Gaussian law
with $\langle\alpha_\mathrm{steep}\rangle = 1.18$ and $\langle\alpha_\mathrm{flat}\rangle =
0.16$ respectively, and a variance of $\sigma_\mathrm{steep,flat}=0.3$.

IR sources are simulated on the basis of a compilation of sources observed by
IRAS. Fluxes are extrapolated to useful frequencies using modified blackbody
spectra $\nu^b B(\nu,T)$, $B(\nu,T)$ being the blackbody function.  Again,
randomly distributed sources are added until the mean surface density as a
function of flux matches everywhere the mean of well covered regions.

The $864$ sources identified as ultra-compact H{\sc ii} regions are treated
separately from the above two populations. Their measured fluxes in the IRAS
catalogues at $100$ and $60 \, \mu m$, together with counterparts in the radio
domain, are fitted with the sum of a modified blackbody and a free-free
spectrum.

Finally, the PSM simulations include a background of unresolved high redshift
faint sources (dusty and star forming galaxies), which is used to model
anisotropies of the cosmic infrared background.

When all these components have been computed for all nine Planck frequencies,
all observed component maps are then coadded to have a full-sky foreground map
which is then added to our CMB and instrument noise simulations.

\subsubsection{Masking}
One of the difficulties with analyzing experimental data is that it is not
possible to use the entire sky for the analysis, because experimental data may
not cover the full sky in every channel, and because the brightest parts of
the sky need to be masked to produce reliable results.  Our method, if it is
to be applicable to experimental data, needs to take this into account.  This
lack of full sky coverage breaks the assumption of independence of the
$a_{\ell m}$ values.  Therefore, we take the strategy of cutting as little of
the sky as possible and hope that the assumption of independence is broken
little enough that the analysis remains robust.

To this end, the masking strategy we use is to take 1.2\% of the brightest
pixels in each frequency map to make a mask at each frequency, then combine
these masks to ensure any pixel masked in any frequency is masked.  Point
sources are not treated separately, though many of the brighter sources are
masked through this method.  In total, this cuts about 1.6\% of the sky.
Then, to prevent aliasing, we apodize the mask with a 33' width apodization
filter (33' is the resolution of the 30 GHz channel).  The apodization filter
we use sets the value of any unmasked pixel equal to
$e^{-\delta\theta^2/2\sigma^2}$ where $\delta\theta$ is the angular distance
to the nearest masked pixel.  This apodization filter ensures that all masked
pixels remain completely masked, and both masked point sources and larger
masked regions have similar apodization applied.  For computational speed, any
unmasked pixel that would have a value greater than 0.99 in the filter is left
untouched.  After applying the apodization, about 3.3\% of the sky is removed
in total.  The sky fraction, $f_{\mathrm{sky}}=0.967$, is estimated using the
average of the squares of the mask pixels, because the $a_{\ell m}$ values are
squared when summing the power spectrum.

The particular masking level is chosen to ensure that the mask is as small as
possible, to prevent correlations, while still not producing obviously wrong
values for the low-$\ell$ power spectrum, as smaller masks tend to produce
aliasing of the foreground signal into the low multipoles, resulting in one or
more low-$\ell$ values being a couple of orders of magnitude larger than
expected.  Additionally, even slightly larger initial masks result in removing
many more point sources which, after apodization, results in much larger sky
fraction being removed (e.g. 7.1\% removed in total with an initial mask size
set to 1.5\%).

Even when retaining a large fraction of sky, we risk biasing the estimated
power spectrum by a few percent if we do not take the fractional coverage of
sky into account.  In order to make progress here, we write down the effect of
the mask in harmonic space as follows:
\begin{equation}
\langle C_\ell^\mathrm{masked}\rangle = {1 \over 2\ell+1} \sum_{m=-\ell}^\ell
\left(a_{\ell m} \odot w_{\ell m}\right)\left(a_{\ell m} \odot w_{\ell
  m}\right)^*.
\end{equation}

Here the operator $\odot$ represents a convolution, which can be performed by
converting the spherical harmonic transform of the mask $w_{\ell m}$ and the
spherical harmonic transform of the CMB $a_{\ell m}$ back to pixel space and
then multiplying the two together pixel by pixel.  In practice, this process
will mix the signals betwen different $m$ and $\ell$ values together.
However, if we ignore the correlations induced by this process and assume
statistical isotropy of the CMB, then we can represent the effect of the mask
as a simple rescaling of the power spectrum:
\begin{equation}
\langle C_\ell^\mathrm{masked}\rangle = {1 \over
  N_p} \sum_{i=0}^{N_p-1} w_i^2 {1 \over 2\ell+1} \sum_{m=-\ell}^\ell a_{\ell
  m} a_{\ell m}^*.
\end{equation}

Here $w_i$ represents the value of the mask at pixel $i$, with $w_i = 1$ for
pixels that are fully unmasked, $w_i = 0$ for pixels fully masked, and values
in between for pixels within the apodized region at the borders of the mask.
We can thus represent this sum over the pixels in the mask as a single
parameter:
\begin{equation}
f_{\mathrm{sky}} = {1 \over N_p} \sum_{i=0}^{N_p-1} w_i^2.
\end{equation}

With this definition, we can write our final estimate of the probability
distribution of $\tilde{C}_\ell$ as follows:
\begin{equation}
\langle \tilde{C}_\ell \rangle f_{\mathrm{sky}} = \left(1 + {1 - n_c \over
  n_{\mathrm{eff}}}\right)\langle C_\ell \rangle + {1 \over \bld{e}^\dagger
  \hat{\mathbf{N}}_\ell^{-1} \bld{e}},
  \label{eqn:psky_bias}
\end{equation}
\begin{equation}
\langle \delta\tilde{C}_\ell^2 \rangle f_{\mathrm{sky}}^2 = {2 \langle C_\ell \rangle \over
  n_{\mathrm{eff}}}\left(\left(1 + {1 - n_c \over n_{\mathrm{eff}}}\right)\langle C_\ell
\rangle + {2 \over \bld{e}^\dagger \hat{\mathbf{N}}_\ell^{-1} \bld{e}}\right),
\end{equation}
where $n_{\mathrm{eff}} = (2\ell+1)f_{\mathrm{sky}}$. The motivation for this
change is that the primary effect of the mask is to apply a weight function to
the $a_{\ell m}$ values so that instead of sums over $2\ell+1$ independent
random variables, we have sums over $(2\ell+1)f_{\mathrm{sky}}$ independent
random variables.

While the induced correlations between the $\tilde{C}_\ell$ values are not
taken into account, it is our hope that keeping the masked region as small as
possible minimizes the impact of this choice.  In the following subsection we
go over the tests we have used to ensure that at this level, the likelihood
function is good enough to accurately capture the power spectrum and its
uncertainties.

\begin{figure*}
\begin{center}
\includegraphics[width=0.49\textwidth]{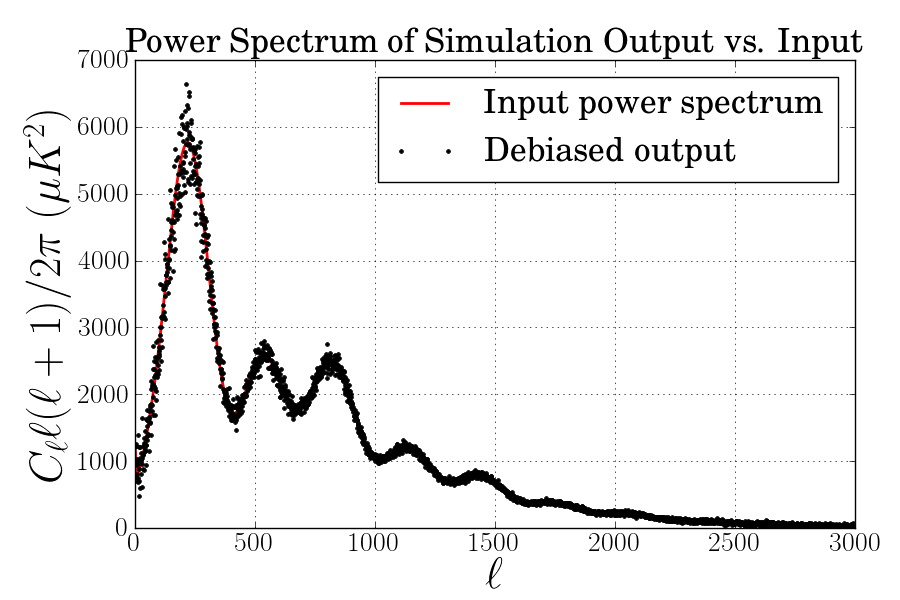}
\includegraphics[width=0.49\textwidth]{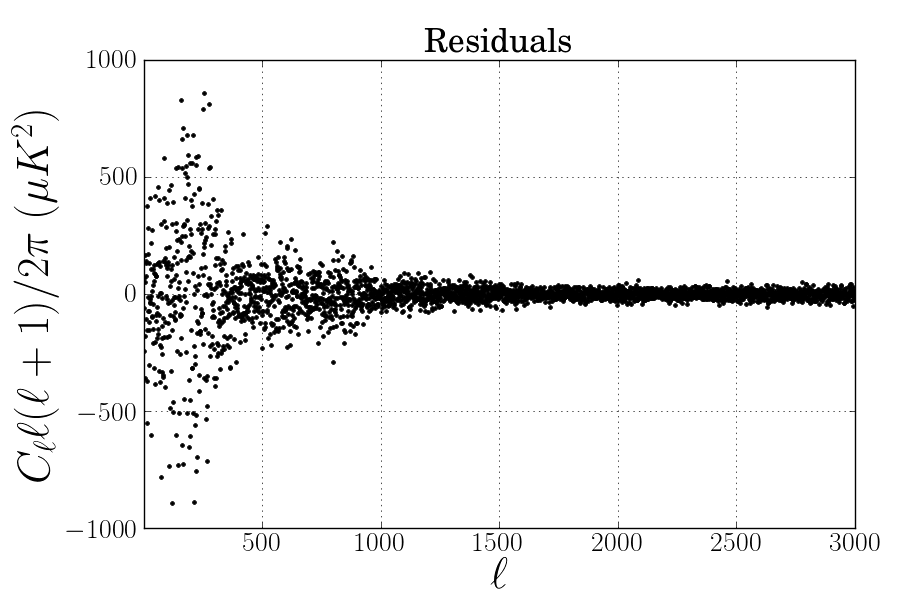}
\end{center}
\caption{These figures compare the power spectra of the simulation input and
  of the component-separated results.  Note that the debiasing which was
  performed assumed perfect knowledge of the foreground plus noise signal as
  well as input CMB, which significantly reduces the scatter, particularly at
  high $\ell$, compared to what would be available in an actual experiment.
  However, these plots are useful to visually demonstrate the validity of
  Equation (\ref{eqn:psky_bias}) when applied to a realistic simulation.  When
  used in an actual experiment, this equation would require factoring in the
  uncertainties in the foreground plus noise signal in some other manner.  The
  primary conclusion here is that the mask, which removes 3.7\% of the sky, does
  not impact the results in a manner which is clearly visible at the
  power spectrum level.}
\label{fig:ps}
\end{figure*}

\subsubsection{The Kolmogorov-Smirnov Test}
In order to determine whether or not our approximation to the likelihood
function is accurate, we make use of a Kolmogorov-Smirnov test, using the
different $\ell$ values as different samples in the test.  The K-S test
compares the empirical cumulative probability distribution of a sample
distribution with the theoretical cumulative probability distribution, and
provides a statistic which can be used to evaluate whether or not the two
distributions are the same.

The Kolmogorov-Smirnov statistic is the area that lies between the cumulative
probability distributions of a test distribution and an empirical
distribution.  If the empirical distribution and the test distribution have
the same shape, then this K-S statistic converges to zero.  In fact, the
convergence to zero occurs at a very specific rate depending upon the number
of samples, if the two probability distributions are the same.

Correlations that are not accounted for in the test distribution will tend to
reduce the rate at which the K-S statistic will converge to zero.  To take a
simple example, consider a situation where the test distribution assumes the
samples are perfectly uncorrelated, while the input data sets every other
value equal to the previous.  With each value in the input data appearing
twice, the K-S statistic will converge half as quickly, leading to a larger
than expected K-S statistic.

Alternatively, if the input data have unmodeled anti-correlations, the reverse
occurs and the K-S statistic converges more rapidly to zero than expected.  A
toy example here is to consider the situation where the test and empirical
distributions both produce numbers between zero and one, but the empirical
distribution sets every other number equal to one minus the previous.  In this
case, the K-S statistic will have encountered all of the independent numbers
while iterating over only half of the range, which in practice speeds its
convergence.

Thus by combining a $\chi^2$ test with the K-S statistic, we can obtain an
estimate of both the shape of the distribution, which the $\chi^2$ test is
sensitive to, and the correlations.

However, we cannot apply the K-S statistic directly with only one realization
because the probability distribution of $\tilde{C}_\ell$ depends upon $\ell$.
We can, however, perform a transformation on the random variable
$\tilde{C}_\ell$ into some target probability distribution which is
easily-estimated.  A unit-variance Gaussian distribution is ideal here.  Such
transformation between different probability distributions is, for instance,
routine in the generation of pseudorandom numbers.  This is done by equating
the cumulative probability distributions:
\begin{equation}
{1 \over \sqrt{2\pi}} \int_{-\infty}^{\chi_\ell} e^{x^2 \over 2}dx =
  \int_{-\infty}^{\tilde{C}_\ell} P(\tilde{C}_\ell') d\tilde{C}_\ell'.
\end{equation}
Note that for pseudorandom number generation, the source distribution is
usually the uniform distribution on the interval $[0,1)$, for which the
cumulative distribution is simply the random number itself.  This sometimes
masks the fact that what is being done is the equating of two cumulative
distributions, making it appear as if one cumulative distribution is being
equated to a random number.  Our situation is not quite that simple.

Since the left hand side is the cumulative distribution of a unit-variance
Gaussian distribution in one variable, we know the $\chi^2_\ell$ will follow a
$\chi^2$ distribution with one degree of freedom.  We convert to a Gaussian
distribution first and then to a $\chi^2$ distribution because it is much
simpler to invert the cumulative Gaussian distribution.  The astute reader may
note that the left hand side of the equation is simply related to the error
function, and that we can compute $\chi^2_\ell$ using the inverse of the error
function as follows:
\begin{equation}
\chi^2_\ell = 2 \left(\mathrm{erf}^{-1}\left(2\int_{-\infty}^{\tilde{C}_\ell}
P(\tilde{C}_\ell') d\tilde{C}_\ell'-1\right)\right)^2.
\end{equation}

If our approximation to the full probability distribution of $\tilde{C}_\ell$
is sufficiently accurate, then the $\chi^2_\ell$ for each $\ell$ follows a
$\chi^2$ distribution with one degree of freedom.  That is,
\begin{equation}
P(\chi^2_\ell) = {e^{-{\chi_\ell}^2 \over 2} \over \sqrt{2\pi \chi_\ell^2}}.
\end{equation}
For the inverse of the error function, we make use of an approximation with
elementary functions produced by Sergei
Winitzki\footnote{\scriptsize{\url{https://sites.google.com/site/winitzki/sergei-winitzkis-files/erf-approx.pdf}}},
using methodology described in \citet{winitzki03},
\begin{eqnarray}
\mathrm{erf}^{-1}(x) &\approx& \mathrm{sgn}(x)\sqrt{\sqrt{\left(f(x)\right)^2 - {\ln (1-x^2)
      \over a}} - f(x)},\\
f(x) &\equiv& {2 \over \pi a} + {\ln (1-x^2) \over 2},
\end{eqnarray}
using the value of $a=0.147$ to keep the relative error at around the
$10^{-4}$ level for all values of the argument of the inverse error function.

Note that while there are more computationally efficient methods to convert
from some given probability distribution to a unit-variance Gaussian, they
typically involve tricks such as combining two random numbers in non-trivial
ways, or throwing away some random numbers.  Neither of which would be an
ideal solution here, as we only have a limited number of $\tilde{C}_\ell$
numbers to transform, and mixing multiple $\tilde{C}_\ell$ values would make
the outcome for correlated or anti-correlated $\tilde{C}_\ell$ values
difficult to predict.  By using this direct transformation, the transformation
reduces to what accounts to mostly a rescaling of the distribution of
$\tilde{C}_\ell$ (as $\chi$ is monotonically increasing along with
$\tilde{C}_\ell$).  By just rescaling the distribution, we are likely to
retain any correlations that exist, especially at higher $\ell$ where the
rescaling is nearly uniform as $P(\tilde{C}_\ell)$ approaches a Gaussian
distribution.

Finally, in order to determine whether or not our computed value of the K-S
statistic is reasonable, we compare this K-S statistic to a set of $10^4$
realizations of 2999 $\chi^2$-distributed independent random variables,
quantifying the result using the fraction which provide a K-S statistic
greater than the K-S statistic of our simulation result.

The results of these tests are that at the level of a single realization of
the sky and over an $\ell$ range from $2-3000$, both the K-S statistic and the
$\chi^2$ statistic lie within the 68\% confidence limits.  These results are
shown in table \ref{tbl:stat_test}.

\begin{table}
  \begin{center}
    \begin{tabular}{ r | c | c}
      \multicolumn{3}{|c|}{$\chi^2$ test}\\
      \hline
      & $\chi^2$ & expected $\chi^2$\\
      \hline
      Analytical & 2952 & 2999 $\pm$ 77.4\\
      Gaussian & 2956 & 2999 $\pm$ 77.4\\
      \hline
      \multicolumn{3}{|c|}{K-S test}\\
      \hline
      & K-S statistic (D$_\mathrm{a}$) & P(D $>$ D$_\mathrm{a}$)\\
      Analytical & 0.0138 & 0.60 \\
      Gaussian & 0.0133 & 0.65 \\
      \hline
    \end{tabular}
    \label{tbl:stat_test}
    \caption{The results of these statistical tests show that the probability
      distributions of the resultant $\tilde{C}_\ell$ values are in good
      accordance with the expected distributions to within the sample variance
      given by the $\ell$ range used.  The K-S statistic found is within the
      one-sigma significance for both the Analytical and Gaussian estimates,
      indicating no significant unaccounted for net correlations or
      anti-correlations in the data.  The expected error on the $\chi^2$ test
      comes from the variance of the $\chi^2$ distribution, which is twice the
      number of degrees of freedom.  The probability P(D $>$ D$_\mathrm{a}$)
      is the probability of obtaining a higher K-S statistic, as estimated by
      $10^4$ simulations of sets of 2999 $\chi^2$- distributed random
      variables.  Values close to $0.5$ are expected.}
  \end{center}
\end{table}

\section{Discussion}
\label{sec:discussion}
\begin{figure}
\begin{center}
\includegraphics[width=0.45\textwidth]{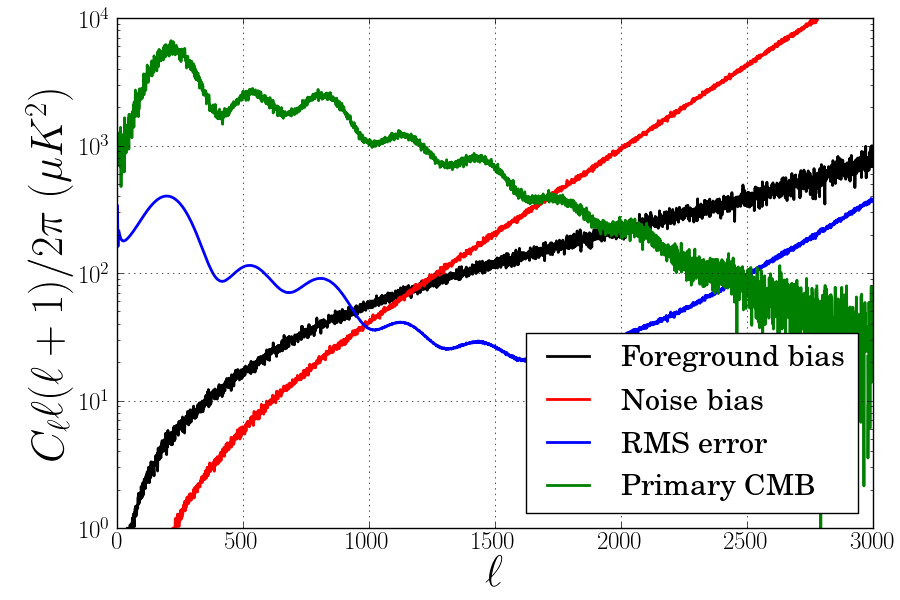}
\end{center}
\caption{Comparison of the foreground and noise bias components against the
  debiased CMB power spectrum and an estimate of the RMS error.  This estimate
  of the RMS error is based upon Equation (\ref{eqn:mod_var}), which adds a
  very rough estimate of the error due to the uncertainties in the foregrounds
  plus noise, assuming the foregrounds plus noise have a known power spectrum
  and are Gaussian-distributed.  Because the foregrounds are negligible at low
  $\ell$ compared to cosmic variance, an accurate diffuse foreground model is
  probably not necessary.  With the foregrounds becoming significant compared
  to the overall uncertainty at around $\ell > 900$, an accurate compact
  source model is required, as the level of the foreground residuals becomes
  larger than the total expected error in the power spectrum.  This supports
  the use of PSM-like simulations for diffuse foregrounds, with the
  uncertainties ignored, combined with a physical model for the compact
  foregrounds whose parameters are estimated along with the cosmological
  parameters.}
\label{fig:fg_ps}
\end{figure}

In this paper, we have presented a new likelihood function which can be used to
estimate the likelihood of the CMB power spectrum given the data and a
foreground model.  Of particular interest is that with the simulations used in
this paper, Planck data shows very little dependence on the foreground model
until $\ell > 2000$ or so, as shown in Figure \ref{fig:fg_ps}.  If true, this
would be an indication that Planck has adequate frequency coverage for the
level of diffuse foregrounds in temperature data.

For $\ell > 1000$ or so, however, it is notable that in order to match the
simulations, we need not only a model of instrument noise, but also the
foreground model.  This is an indication that compact sources remain a
significant factor in the data.  From these tests, it would be perfectly
reasonable to make use of a fixed Galactic foreground model combined with a
power spectrum-based compact source model, such as that used in
\citet{millea11}.  This is very much an ideal situation because the parameters
that impact the power spectra of compact sources can be very well-estimated at
the power spectrum level.

The most accurate way of including the instrument noise contribution would
probably be to pre-compute the spherical harmonic transforms of a set of
simulations of the full instrument noise.  While the number of noise
simulations is likely to be much smaller than the number of Monte Carlo
simulations in the full parameter estimation analysis, a thousand or so noise
simulations should be sufficient to effectively marginalize over instrument
noise.  This method would have the advantage of incorporating the full
instrument noise covariance, if sufficiently-detailed noise simulations are
used, albeit in an approximate manner.  Care would need to be taken to ensure
that the use of the simulations within the parameter-estimation MCMC chain do
not require significant computing resources.  One might also imagine simpler
approximations such as that used in Equation (\ref{eqn:mod_var}).

Therefore we would recommend making use of the probability function
$P(\tilde{C}_\ell|C_\ell,N_\ell)$, where $N_\ell = 1/e^\dagger
\hat{\mathbf{N}}_\ell^{-1} e$, with $\hat{\mathbf{N}}_\ell$ being the
estimated covariance of the instrument noise plus foreground model which
depends upon any foreground parameters which are to be marginalized over.

Once the details of the foreground plus noise models are determined, the work
in this paper demonstrates that provided the mask removes very little of the
sky, the likelihood function we supply provides a reasonably accurate estimate
of the true likelihood of the CMB in the presence of foregrounds.  However, it
does have difficulties at low multipoles, as can be seen through a cursory
examination of Equation (\ref{eqn:bias}).  In particular the second term,
$(1-n_c)/(2\ell+1) \langle C_\ell \rangle$, represents the statistical
component of the ILC bias.  Note that the denominator is the number of
$a_{\ell m}$ terms that go into the average for this power spectrum component.

Perhaps a useful comparison of how significant this bias term is would be to
compare it against the fact that other power spectrum estimation codes use a
much smaller fraction of the sky in their analysis with a similar impact on
the error that results.  If we are to compare against a method which uses 80\%
sky coverage, for instance, a back-of-the-envelope calculation would suggest
that we want our statistical bias term to be no more than about 15\% or so of
the power spectrum, indicating we should have a better estimate of the errors
for roughly $\ell > 50$ as compared with such a hypothetical method.  While
some may be nervous about trusting our estimate of the errors in the
intermediate $\ell$ range of around $50 < \ell < 400$ or so, our simulations
indicate that the residual foreground contamination in this range is more than
an order of magnitude below the estimated total error in the power spectrum.

Additionally, it might be worthwhile to consider binning the power spectrum in
order to reduce the magnitude of the statistical bias term.  However, binning
the power spectrum does destroy some information about the CMB power, and it
may potentially increase the foreground bias term, so we do not consider that
here.  One additional possibility is a hybrid method: estimate the ILC weights
on bins in $\ell$, but extract the power spectrum separately at each $\ell$.
This has the benefit of reducing the statistical bias term while retaining
full resolution on the CMB power.  However, it has the drawback that it likely
increases the foreground bias term, while at the same time inducing
correlations between each $\tilde{C}_\ell$ within a bin.

While it is in principle possible to estimate the correlation within each bin,
at very low-$\ell$ the likelihood function is non-Gaussian, and it is an
extremely non-trivial problem to consider both the correlations and the
non-Gaussianity of the low-$\ell$ power spectrum at the same time.  There is,
however, an intermediate range in $\ell$, around $\ell=100$ to $\ell=400$ or
so, where the statistical bias term is non-negligible but the likelihood
function highly Gaussian where this approach may be tractable.  But due to the
work involved for what is likely to be a very small improvement in the CMB
uncertainty (of the order of a few percent in the variance in
$\tilde{C}_\ell$), we do not consider this approach here.

Another potential question is why we do not suggest simply making use of a
model of the foregrounds given by the data minus the estimated CMB.  As we
show in Appendix \ref{app:bootstrap}, using this as the foreground model gives
identically zero for the estimate of the foreground contamination.  In fact,
this calculation places into doubt any sort of ``boot strapping''-type
estimation of the foreground subtraction errors: it is absolutely required to
have some physical model of the foregrounds in order to estimate the
foreground subtraction errors.

This paper also begs the question that if we require a foreground model such
as that used in \citet{millea11}, why go through the work of implementing the
ILC at all?  Why not simply make use of the parametric model of the
foregrounds?  Our argument here is that because the ILC method significantly
reduces the foregrounds prior to the likelihood analysis, this method will be
less sensitive to the precise details of the foreground model used, permitting
a greater level of uncertainty in the foreground model before cosmological
parameters are impacted.  This is particularly the case at high $\ell$ when we
are no longer limited by cosmic variance.

\section*{Acknowledgments}

Some of the results in this paper have been derived using the HEALPix package
\citep{gorski05}, CAMB \citep{lewis00}, and LensPix \citep{lewis05}.  We would
also like to thank Samuel Leach and Carlo Baccigalupi for some useful
conversations.  Guillaume Castex would like to thank SISSA for the hospitality
during this work.

\bibliography{./cmb4}

\appendix
\onecolumn

\section{Variance Calculation}
\label{app:variance}

In order to compute the variance of the estimated power spectrum, we start
with Equation (\ref{eqn:variance}), the square of which is:

\begin{eqnarray}
\delta\tilde{C}_\ell^2 &=& \nonumber \\
 && \delta C_\ell^2 + \left({1 - n_c \over
  2\ell+1}\langle C_\ell \rangle \right)^2 + \left(\bld{f}^\dagger_\ell
  \hat{\mathbf{N}}_\ell^{-1} \bld{f}_\ell\right)^2
  + \left({\bld{e}^\dagger \hat{\mathbf{N}}_\ell^{-1}
  \bld{f}_\ell + \bld{f}^\dagger_\ell \hat{\mathbf{N}}_\ell^{-1}
  \bld{e} \over \bld{e}^\dagger \hat{\mathbf{N}}_\ell^{-1}
  \bld{e}}\right)^2 
  + \left({\bld{e}^\dagger \hat{\mathbf{N}}_\ell^{-1}
  \bld{f}_\ell \bld{f}^\dagger_\ell \hat{\mathbf{N}}_\ell^{-1}
  \bld{e} \over \bld{e}^\dagger \hat{\mathbf{N}}_\ell^{-1}
  \bld{e}}\right)^2 \nonumber \\
 && + 2\delta C_\ell \left({1 - n_c \over
  2\ell+1}\langle C_\ell \rangle - \bld{f}^\dagger_\ell
  \hat{\mathbf{N}}_\ell^{-1} \bld{f}_\ell + {\bld{e}^\dagger
  \hat{\mathbf{N}}_\ell^{-1} \bld{f}_\ell + \bld{f}^\dagger_\ell
  \hat{\mathbf{N}}_\ell^{-1} \bld{e} + \bld{e}^\dagger \hat{\mathbf{N}}_\ell^{-1}
  \bld{f}_\ell \bld{f}^\dagger_\ell \hat{\mathbf{N}}_\ell^{-1}
  \bld{e} \over \bld{e}^\dagger \hat{\mathbf{N}}_\ell^{-1}
  \bld{e}}\right) \nonumber \\
 && + 2 \left({1 - n_c \over 2\ell+1}\right)\langle C_\ell \rangle \left(- \bld{f}^\dagger_\ell
  \hat{\mathbf{N}}_\ell^{-1} \bld{f}_\ell + {\bld{e}^\dagger
  \hat{\mathbf{N}}_\ell^{-1} \bld{f}_\ell + \bld{f}^\dagger_\ell
  \hat{\mathbf{N}}_\ell^{-1} \bld{e} + \bld{e}^\dagger \hat{\mathbf{N}}_\ell^{-1}
  \bld{f}_\ell \bld{f}^\dagger_\ell \hat{\mathbf{N}}_\ell^{-1}
  \bld{e} \over \bld{e}^\dagger \hat{\mathbf{N}}_\ell^{-1}
  \bld{e}}\right) \nonumber \\
 && - 2 \bld{f}^\dagger_\ell \hat{\mathbf{N}}_\ell^{-1} \bld{f}_\ell \left(
  {\bld{e}^\dagger \hat{\mathbf{N}}_\ell^{-1} \bld{f}_\ell + \bld{f}^\dagger_\ell
  \hat{\mathbf{N}}_\ell^{-1} \bld{e} + \bld{e}^\dagger \hat{\mathbf{N}}_\ell^{-1}
  \bld{f}_\ell \bld{f}^\dagger_\ell \hat{\mathbf{N}}_\ell^{-1}
  \bld{e} \over \bld{e}^\dagger \hat{\mathbf{N}}_\ell^{-1}
  \bld{e}}\right) \nonumber \\
 && + 2 \left({\bld{e}^\dagger \hat{\mathbf{N}}_\ell^{-1}
  \bld{f}_\ell + \bld{f}^\dagger_\ell \hat{\mathbf{N}}_\ell^{-1}
  \bld{e} \over \bld{e}^\dagger \hat{\mathbf{N}}_\ell^{-1}
  \bld{e}}\right) \left({\bld{e}^\dagger \hat{\mathbf{N}}_\ell^{-1}
  \bld{f}_\ell \bld{f}^\dagger_\ell \hat{\mathbf{N}}_\ell^{-1}
  \bld{e} \over \bld{e}^\dagger \hat{\mathbf{N}}_\ell^{-1}
  \bld{e}}\right).
\end{eqnarray}
This equation seems quite daunting, however when we take the expectation
value, many of the terms turn out to be zero.  The first aspect which we
exploit is that only even products of the $a_{\ell m}$ coefficients have
non-zero expectation value.  Note that $\bld{f}_\ell$ is linear in $a_{\ell
  m}$, while $C_\ell$ goes as the square of $a_{\ell m}$.  Additionally,
because $\hat{\mathbf{N}}_\ell$ is a real, symmetric matrix, $\bld{e}^\dagger
\hat{\mathbf{N}}_\ell^{-1} \bld{f}_\ell = \bld{f}^\dagger_\ell
\hat{\mathbf{N}}_\ell^{-1} \bld{e}$.  Combining these two statements along
with $\langle \delta C_\ell \rangle = 0$ gives:
\begin{eqnarray}
\langle\delta\tilde{C}_\ell^2\rangle &=& \nonumber \\
 && \left\langle\delta C_\ell^2 + \left({1 - n_c \over
  2\ell+1}\langle C_\ell \rangle \right)^2 + \left(\bld{f}^\dagger_\ell
  \hat{\mathbf{N}}_\ell^{-1} \bld{f}_\ell\right)^2
  + \left({2\bld{e}^\dagger \hat{\mathbf{N}}_\ell^{-1}
  \bld{f}_\ell \over \bld{e}^\dagger \hat{\mathbf{N}}_\ell^{-1}
  \bld{e}}\right)^2
  + {\left(\bld{e}^\dagger \hat{\mathbf{N}}_\ell^{-1}
  \bld{f}_\ell\right)^4 \over \left(\bld{e}^\dagger \hat{\mathbf{N}}_\ell^{-1}
  \bld{e}\right)^2}\right. \nonumber \\
 && + 2\delta C_\ell \left( - \bld{f}^\dagger_\ell
  \hat{\mathbf{N}}_\ell^{-1} \bld{f}_\ell + {\left(\bld{e}^\dagger
  \hat{\mathbf{N}}_\ell^{-1} \bld{f}_\ell\right)^2 \over \bld{e}^\dagger
  \hat{\mathbf{N}}_\ell^{-1} \bld{e}}\right) 
 + 2 \left({1 - n_c \over 2\ell+1}\right)\langle C_\ell \rangle \left(- \bld{f}^\dagger_\ell
  \hat{\mathbf{N}}_\ell^{-1} \bld{f}_\ell + {\left(\bld{e}^\dagger \hat{\mathbf{N}}_\ell^{-1}
  \bld{f}_\ell\right)^2 \over \bld{e}^\dagger \hat{\mathbf{N}}_\ell^{-1}
  \bld{e}}\right) \nonumber\\
 &&- 2 \left.\bld{f}^\dagger_\ell \hat{\mathbf{N}}_\ell^{-1} \bld{f}_\ell 
  {\left(\bld{e}^\dagger \hat{\mathbf{N}}_\ell^{-1}
  \bld{f}_\ell\right)^2 \over \bld{e}^\dagger \hat{\mathbf{N}}_\ell^{-1}
  \bld{e}}\right\rangle.
\label{eqn:full_var}
\end{eqnarray}

From here, we need to individually compute the expectation values of each
component.  This is done using the known expectation values of the $a_{\ell
  m}$ coefficients:
\begin{eqnarray}
\langle a_{\ell m_1}a^*_{\ell m_2} \rangle &=& \langle C_\ell \rangle \delta_{m_1 m_2},\\
\langle a_{\ell m_1}a^*_{\ell m_2}a_{\ell m_3}a^*_{\ell m_4} \rangle &=& \langle
C_\ell \rangle^2 \left( \delta_{m_1 m_2}\delta_{m_3 m_4} + \delta_{m_1
  m_4}\delta_{m_2 m_3} + \delta_{m_1 -m_3}\delta_{m_2 -m_4}\right).
\label{eqn:alm_kurt}
\end{eqnarray}
Here we describe in detail the calculation of one of the more complicated
terms, and simply list the answers for the remaining terms.  Perhaps the most
complicated term is the final one, which requires taking the expectation value
of the following:
\begin{equation}
\left\langle { \bld{f}^\dagger_\ell \hat{\mathbf{N}}_\ell^{-1} \bld{f}_\ell 
  \left(\bld{e}^\dagger \hat{\mathbf{N}}_\ell^{-1}
  \bld{f}_\ell\right)^2 \over \bld{e}^\dagger \hat{\mathbf{N}}_\ell^{-1}
  \bld{e}} \right\rangle = \left\langle { \bld{f}^\dagger_\ell \hat{\mathbf{N}}_\ell^{-1} \bld{f}_\ell 
  \bld{f}^\dagger_\ell \hat{\mathbf{N}}_\ell^{-1}
  \bld{e}\bld{e}^\dagger \hat{\mathbf{N}}_\ell^{-1}
  \bld{f}_\ell \over \bld{e}^\dagger \hat{\mathbf{N}}_\ell^{-1}
  \bld{e}} \right\rangle.
  \label{eqn:expec_example}
\end{equation}
We expand the square in this particular way in order to ensure that when expanded
fully, the $a_{\ell m}$ coefficients would have the same order of complex
conjugation as in Equation (\ref{eqn:alm_kurt}).  The full expansion is done by
using the definition of $\bld{f}_\ell$ in Equation (\ref{eqn:fl_def}).  Note
that the conjugate transpose of $\bld{f}_\ell$ is then defined as follows:
\begin{equation}
\bld{f}_\ell^\dagger \equiv \sum_{m = -\ell}^\ell a_{\ell m}\bld{f}_{\ell m}^\dagger.
\end{equation}
With these definitions, Equation (\ref{eqn:expec_example}) becomes:
\begin{eqnarray}
\left\langle { \bld{f}^\dagger_\ell \hat{\mathbf{N}}_\ell^{-1} \bld{f}_\ell 
  \bld{f}^\dagger_\ell \hat{\mathbf{N}}_\ell^{-1}
  \bld{e}\bld{e}^\dagger \hat{\mathbf{N}}_\ell^{-1}
  \bld{f}_\ell \over \bld{e}^\dagger \hat{\mathbf{N}}_\ell^{-1}
  \bld{e}} \right\rangle &=& \nonumber\\{1 \over (2\ell+1)^4} &&\sum_{m_1 =
  -\ell}^\ell\sum_{m_2 = -\ell}^\ell\sum_{m_3 = -\ell}^\ell\sum_{m_4 =
  -\ell}^\ell \langle a_{\ell m_1}a^*_{\ell m_2}a_{\ell m_3}a^*_{\ell m_4}
\rangle {\bld{f}_{\ell m_1}^\dagger \hat{\mathbf{N}}_\ell^{-1}
  \bld{f}_{\ell m_2}\bld{f}_{\ell m_3}^\dagger \hat{\mathbf{N}}_\ell^{-1}
  \bld{e}\bld{e}^\dagger \hat{\mathbf{N}}_\ell^{-1}
  \bld{f}_{\ell m_4} \over \bld{e}^\dagger \hat{\mathbf{N}}_\ell^{-1}
  \bld{e}}.
\end{eqnarray}
We can then make use of the definition of $\hat{\mathbf{N}}_\ell$, which is:
\begin{equation}
\hat{\mathbf{N}}_\ell \equiv {1 \over 2\ell+1} \sum_{m = -\ell}^\ell
\bld{f}_{\ell m} \bld{f}_{\ell m}^\dagger,
\end{equation}
as well as Equation (\ref{eqn:alm_kurt}) and the fact that $\bld{f}_{\ell m}^* =
\bld{f}_{\ell -m}$ to give:
\begin{equation}
\left\langle { \bld{f}^\dagger_\ell \hat{\mathbf{N}}_\ell^{-1} \bld{f}_\ell 
  \bld{f}^\dagger_\ell \hat{\mathbf{N}}_\ell^{-1}
  \bld{e}\bld{e}^\dagger \hat{\mathbf{N}}_\ell^{-1}
  \bld{f}_\ell \over \bld{e}^\dagger \hat{\mathbf{N}}_\ell^{-1}
  \bld{e}} \right\rangle = {\langle C_\ell \rangle^2 \over (2\ell+1)^2}
  \left({\mathbf{Tr[\hat{\mathbf{N}}_\ell^{-1}\hat{\mathbf{N}}_\ell] \bld{e}^\dagger \hat{\mathbf{N}}_\ell^{-1}
  \hat{\mathbf{N}}_\ell \hat{\mathbf{N}}_\ell^{-1}
  \bld{e}} \over \bld{e}^\dagger \hat{\mathbf{N}}_\ell^{-1}
  \bld{e}}
 + 2{\bld{e}^\dagger \hat{\mathbf{N}}_\ell^{-1}
  \hat{\mathbf{N}}_\ell\hat{\mathbf{N}}_\ell^{-1}
  \hat{\mathbf{N}}_\ell\hat{\mathbf{N}}_\ell^{-1}
  \bld{e} \over \bld{e}^\dagger \hat{\mathbf{N}}_\ell^{-1}
  \bld{e}}
 \right).
\end{equation}

Since the trace of the identity matrix is simply the number of channels, this
expectation value simplifies to:
\begin{equation}
\left\langle { \bld{f}^\dagger_\ell \hat{\mathbf{N}}_\ell^{-1} \bld{f}_\ell 
  \bld{f}^\dagger_\ell \hat{\mathbf{N}}_\ell^{-1}
  \bld{e}\bld{e}^\dagger \hat{\mathbf{N}}_\ell^{-1}
  \bld{f}_\ell \over \bld{e}^\dagger \hat{\mathbf{N}}_\ell^{-1}
  \bld{e}} \right\rangle = {\langle C_\ell \rangle^2 \over (2\ell+1)^2}
\left(n_c + 2\right).
\end{equation}
The expectation values of the remaining terms in Equation (\ref{eqn:full_var})
are:
\begin{eqnarray}
\langle\delta C_\ell^2\rangle &=& {2 \langle C_\ell\rangle^2 \over 2\ell+1},\\
\left\langle\left(\bld{f}^\dagger_\ell \hat{\mathbf{N}}_\ell^{-1}
\bld{f}_\ell\right)^2\right\rangle &=& \langle C_\ell\rangle^2{n_c^2 + 2n_c
  \over (2\ell+1)^2},\\
\left\langle{\bld{e}^\dagger \hat{\mathbf{N}}_\ell^{-1}
\bld{f}_\ell \bld{f}^\dagger_\ell \hat{\mathbf{N}}_\ell^{-1} \bld{e} \over
\left(\bld{e}^\dagger \hat{\mathbf{N}}_\ell^{-1} \bld{e}\right)^2}\right\rangle &=& {\langle
C_\ell \rangle \over (2\ell+1) \bld{e}^\dagger \hat{\mathbf{N}}_\ell^{-1}
\bld{e}},\\
\left\langle\left( {\bld{e}^\dagger \hat{\mathbf{N}}_\ell^{-1}
\bld{f}_\ell \bld{f}^\dagger_\ell \hat{\mathbf{N}}_\ell^{-1} \bld{e} \over
\bld{e}^\dagger \hat{\mathbf{N}}_\ell^{-1} \bld{e}} \right)^2 \right\rangle
&=& {3\langle C_\ell \rangle^2 \over (2\ell + 1)^2},\\
\left\langle \delta C_\ell \bld{f}^\dagger_\ell
\hat{\mathbf{N}}_\ell^{-1}\bld{f}_\ell \right\rangle &=& {2 \langle C_\ell
  \rangle^2 n_c \over (2\ell+1)^2},\\
 \left\langle \delta C_\ell {\bld{e}^\dagger \hat{\mathbf{N}}_\ell^{-1}
\bld{f}_\ell \bld{f}^\dagger_\ell \hat{\mathbf{N}}_\ell^{-1} \bld{e} \over
\bld{e}^\dagger \hat{\mathbf{N}}_\ell^{-1} \bld{e}} \right\rangle &=& {2
   \langle C_\ell \rangle^2 \over (2\ell+1)^2},\\
\left \langle \bld{f}^\dagger_\ell \hat{\mathbf{N}}_\ell^{-1}
\bld{f}_\ell \right\rangle &=& {\langle C_\ell \rangle n_c \over 2\ell+1},\\
\left\langle {\bld{e}^\dagger \hat{\mathbf{N}}_\ell^{-1}
\bld{f}_\ell \bld{f}^\dagger_\ell \hat{\mathbf{N}}_\ell^{-1} \bld{e} \over
\bld{e}^\dagger \hat{\mathbf{N}}_\ell^{-1} \bld{e}} \right\rangle &=& {\langle
  C_\ell \rangle \over 2\ell+1}. 
\end{eqnarray}

Substituting these expectation values into Equation (\ref{eqn:full_var}) gives:
\begin{eqnarray}
\langle\delta \tilde{C}_\ell^2 \rangle =& {\langle C_\ell \rangle \over
  2\ell+1} \left( 2\langle C_\ell \rangle + {(1-n_c)^2\langle C_\ell\rangle \over 2\ell+1} + \langle C_\ell \rangle {n_c^2
  + 2n_c \over 2\ell+1} + {4 \over \bld{e}^\dagger
  \hat{\mathbf{N}}_\ell^{-1} \bld{e}} + {3\langle C_\ell \rangle
  \over 2\ell+1} - {4\langle C_\ell \rangle n_c \over 2\ell+1} + {4\langle
  C_\ell \rangle \over 2\ell+1}\right.\nonumber\\
&\left.- {1-n_c \over 2\ell+1}2\langle C_\ell \rangle
(-n_c + 1) - {2\langle C_\ell \rangle \over 2\ell+1} (n_c + 2)\right),
\end{eqnarray}
which simplifies to:
\begin{equation}
\langle\delta \tilde{C}_\ell^2 \rangle = {2 \langle C_\ell \rangle \over
  2\ell+1}\left(\langle C_\ell \rangle + {1-n_c \over 2\ell+1}\langle C_\ell
\rangle + {2 \over \bld{e}^\dagger \hat{\mathbf{N}}_\ell^{-1} \bld{e}}
\right).
\end{equation}

\section{On the Need for a Physical Model}
\label{app:bootstrap}

There are significant difficulties in developing an accurate physical model
for the foreground signal.  For diffuse Galactic foregrounds, variation along
the line of sight of the foreground properties (such as the dust temperature)
have the potential to significantly complicate the modeling of the foreground.
For compact sources, the spectra of these sources may vary significantly
depending upon the source, and may show some variation over the age of the
universe as well, which would make the process of validating compact source
spectra by using the nearby bright sources a questionable proposition.
Therefore, an absolutely ideal situation would be one in which we are able to
ignore the potential difficulties of finding an accurate model for foreground
emission and are able to extract the primary CMB signal without worrying about
the details of the foregrounds.

This may be one of the motivating factors for the use of so many blind
component separation methods, such as the many variations of ILC and ICA that
have been used.  In fact, it is often impressive how well these component
separation methods do at removing the foregrounds.  However, as we argue is the
case here for ILC in harmonic space, there is no way to estimate the actual
errors in the removal without a physical model.  To do this, we examine
the case where instead of having a physical model for the covariance of the
foregrounds $\hat{\mathbf{N}}_\ell$, we make use of an estimate of the
foreground signal given by:
\begin{equation}
\tilde{\bld{f}}_{\ell m} = \bld{d}_{\ell m} - \tilde{a}_{\ell m}.
\end{equation}
Here $\tilde{\bld{f}}_{\ell m}$ is the spherical harmonic transform of our
estimate of the foreground signal at each observation frequency,
$\bld{d}_{\ell m}$ is the spherical harmonic transform of the data at
each frequency, and $\tilde{a}_{\ell m}$ is the spherical harmonic transform
of our estimate of the CMB sky through harmonic-space ILC.

With these definitions, it is easy to show that the covariance matrix at each
$\ell$ of our estimate of the foreground signal is given by:
\begin{equation}
\tilde{\mathbf{N}}_\ell = \hat{\mathbf{C}}_\ell - \bld{e} \bld{e}^\dagger \tilde{C}_\ell.
\end{equation}

To understand how the use of this estimate of the foreground signal affects
our results, we need to use this estimate for the term that encodes the
effect of the foreground power spectrum in our bias and variance estimates,
$1/e^\dagger\tilde{\mathbf{N}}_\ell^{-1}e$.  To facilitate the calculation of this
statement, we write $\tilde{\mathbf{N}}_\ell$ by expanding
$\hat{\mathbf{C}}_\ell$ in terms of its CMB and foreground components as
follows:
\begin{equation}
\tilde{\mathbf{N}}_\ell = \bld{e} \bld{e}^\dagger (C_\ell -
\tilde{C}_\ell) + \bld{e}\bld{f}_\ell^\dagger +
\bld{f}_\ell\bld{e}^\dagger + \hat{\mathbf{N}}_\ell.
\label{eqn:fg_est_covariance}
\end{equation}
Because of the similarity of Equation (\ref{eqn:fg_est_covariance}) with
(\ref{eqn:covariance_short}), we can simply write down the result as a slight
modification of Equation (\ref{eqn:est_cl}):
\begin{eqnarray}
{1 \over e^\dagger\tilde{\mathbf{N}}_\ell^{-1}e} &=& C_\ell - \tilde{C}_\ell \nonumber\\
&&+ {1 + \bld{e}^\dagger \hat{\mathbf{N}}_\ell^{-1}
  \bld{f}_\ell + \bld{f}^\dagger_\ell \hat{\mathbf{N}}_\ell^{-1} \bld{e}
  + \bld{e}^\dagger \hat{\mathbf{N}}_\ell^{-1} \bld{f}_\ell\bld{f}^\dagger_\ell
  \hat{\mathbf{N}}_\ell^{-1} \bld{e} \over \bld{e}^\dagger
  \hat{\mathbf{N}}_\ell^{-1} \bld{e}} \nonumber\\
 &&- \bld{f}^\dagger_\ell \hat{\mathbf{N}}_\ell^{-1} \bld{f}_\ell\\
 &=& \tilde{C}_\ell - \tilde{C}_\ell = 0.
\label{eqn:est_fg}
\end{eqnarray}
Thus, by using this simplified estimate of the foreground signal, all
information about the foreground contamination, both in the bias and variance,
is destroyed.  While we could, in principle, add something to $\tilde{C}_\ell$
to make it so that $1/e^\dagger\tilde{\mathbf{N}}^+e \neq 0$, the difference
between $1/e^\dagger\tilde{\mathbf{N}}^+e$ and zero will just be a function of
the arbitrary offset and not provide us with any new information about the
foreground signal.

Furthermore, though our calculation is performed on a term which was derived
assuming uncorrelated CMB samples, \citet{delabrouille09} argue that this sort
of calculation can be used as an approximation to the true bias even with
correlated samples, at least within the Needlet framework.  Therefore, to
first order, the effect of the foreground contamination is canceled in the
case of correlated CMB samples.  Higher-order terms arising from the
correlations may cause some information about the bias to be recovered, but
with most of the information destroyed by this method, it is unlikely to be
useful.  This has direct implications upon the bias calculated by the WMAP
team in \citet{hinshaw07}, as their estimation of the bias through simulations
directly mirrors the methodology of the calculation performed in this
Appendix\footnote{Though their paper mentions the MEM model was used for the
  foreground estimate, correspondence with the WMAP team combined with our own
  empirical studies confirm that what was actually done was to simply subtract
  the ILC estimate from the sky maps to produce their estimate of the
  foregrounds.}.  This indicates that the bias used in computing the ILC map
published by the WMAP team likely misses most of the actual ILC bias, and may
have higher-order contributions which are not understood and may, in certain
cases, even have the wrong sign.  Similarly, \citet{kim08} also makes use of
an estimate of the foregrounds as being the data minus their ILC-estimated
CMB, and therefore likely underestimates the bias as well.

This calculation seems to argue that in order to obtain an accurate estimate
of the errors and biases in the ILC method, it is fundamentally required to
have a physical, parametric model of the foregrounds.  It is, in principle,
possible to make use of simulated foregrounds.  However as our likelihood
function in Equation (\ref{eqn:likelihood}) demonstrates, the probability
distribution of $\tilde{C}_\ell$ depends both upon the cosmological model and
upon the foreground model, and it is very difficult to vary both in a
simulation environment.  There is the additional concern that some of the
foregrounds, such as SZ clusters, would also depend upon the cosmological
model in a detailed treatment.

\label{lastpage}

\end{document}